\documentclass[twocolumn,numberedappendix, twocolappendix, apj]{openjournal}
\usepackage[colorlinks=true, linkcolor=blue, citecolor=blue, urlcolor=blue]{hyperref}
\usepackage{graphicx}
\usepackage{amsmath,amssymb}
\usepackage{lastpage}
\usepackage[all]{hypcap}
\usepackage[T1]{fontenc}
\usepackage{float}
\usepackage{subfigure}
\usepackage{afterpage}
\usepackage{xcolor}
\usepackage{mathrsfs}
\usepackage{upgreek}
\usepackage{color}
\usepackage[dvipsnames]{xcolor}
\usepackage{longtable}
\usepackage{threeparttable}
\usepackage{multirow}
\usepackage{placeins}
\usepackage{bm}
\usepackage{fontawesome5}
\usepackage{enumitem}
\usepackage{cool}
\usepackage{float}
\makeatletter
\let\newfloat\newfloat@ltx
\makeatother
\usepackage{algorithm}
\usepackage{algpseudocode}
\usepackage{fontawesome5}
\usepackage{soul}
\usepackage{orcidlink}

\usepackage{fontawesome5}
\newcommand{\github}[1]{\href{#1}{\faGithubSquare}}
\newcommand{\githublink}{\github{https://github.com/harrydesmond/Causality_NSA/}}

\def\app#1#2{%
  \mathrel{%
    \setbox0=\hbox{$#1\sim$}%
    \setbox2=\hbox{%
      \rlap{\hbox{$#1\propto$}}%
      \lower1.1\ht0\box0%
    }%
    \raise0.25\ht2\box2%
  }%
}

\begin{document}
\label{firstpage}

\title[The causal structure of galactic astrophysics]{The causal structure of galactic astrophysics}

\author{Harry Desmond$^{*}$}
\email{$^*$harry.desmond@port.ac.uk}
\affiliation{Institute of Cosmology \& Gravitation, University of Portsmouth, Dennis Sciama Building, Portsmouth, PO1 3FX, UK}
\author{Joseph Ramsey}
\affiliation{Carnegie Mellon University Philosophy Department}

\begin{abstract}
	Data-driven astrophysics currently relies on the detection and characterisation of correlations between objects' properties, which are then used to test physical theories that make predictions for them. This process fails to utilise information in the data that forms a crucial part of the theories' predictions, namely which variables are directly correlated (as opposed to accidentally correlated through others), the directions of these determinations, and the presence or absence of confounders that correlate variables in the dataset but are themselves absent from it. We propose to recover this information through \emph{causal discovery}, a well-developed methodology for inferring the causal structure of datasets that is however almost entirely unknown to astrophysics. We develop a causal discovery algorithm suitable for large astrophysical datasets and illustrate it on $\sim$4.5$\times10^5$ nearby galaxies from the Nasa Sloan Atlas, demonstrating its ability to distinguish physical mechanisms that are degenerate on the basis of correlations alone.
\end{abstract}


\section{Introduction}
\label{sec:intro}

Understanding the physical processes that shape galaxies is a central goal of astrophysics. Empirical progress has traditionally relied on identifying correlations between observed properties, which can then be interpreted in light of theoretical models for galaxy formation and used to constrain them. The advent of large surveys and powerful machine learning techniques has greatly expanded our ability to find such statistical associations, uncovering intricate patterns across high-dimensional parameter spaces. However, correlation alone cannot determine causal influences among variables: which properties are actually responsible for \emph{determining} others, in what \emph{direction} this influence goes, and whether there exist \emph{confounding variables} that are not included in the dataset but influence those that are~\citep{Pearl_2009}.
Achieving this requires \emph{causal discovery}, a methodology widely applied in fields such as genomics, epidemiology and economics, but that has had extremely limited exposure in the physical sciences~\citep{review_1,Glymour2019}.

This paper seeks to develop causal discovery to the point where it can be applied to the entire low-redshift galaxy population,
as a method for adding value to traditional correlation or machine learning analyses. This offers the promise of determining whether the empirical links between physical variables reflect causal pathways (indicating a physical operation of one quantity on another) or merely statistical co-variation (indicating an accidental correlation reflecting a physical law in operation elsewhere). This is precisely the kind of information predicted by physical theories, and hence provides great potential for improving constraints on them.

A possibly complete list of previous applications of causal discovery to astrophysics follows.~\citet{astro_eg_4} apply the Peter--Clark and Fast Causal Inference algorithms to 83 galaxies in an attempt to constrain evolution mechanisms for their central supermassive black holes.~\citet{astro_eg_1,astro_eg_0} addresses the same question with 101 galaxies, using a Bayesian method for estimating the probabilities of all possible causal structures.~\citet{astro_eg_3} estimate a causal model of galaxy formation from semi-analytic models and hydrodynamical simulations, and compare it to non-causal models. (They focus mainly however on \emph{causal inference}---the estimation of causal effects given a causal model---rather than causal discovery, the learning of the causal model in the first place.)
\citet{Davis_TNOs} use Fast Causal Inference to help constrain the formation histories of trans-Neptunian objects.
\citet{Zhang_2025} apply a Fast Causal Inference-like algorithm to learn the interrelations between photometric properties of $\sim$200 galaxy clusters.
Finally,~\citet{astro_eg_5} apply a linear causal discovery model with small conditioning set size and the assumption of Gaussian residuals to $\sim150000$ stars in simulated galaxies to constrain chemodynamical pathways relevant for the Milky Way.
These studies involve samples either too small to be representative of the underlying populations, or, in the latter case, making stringent assumptions about the correlation structure that may not hold in real astrophysical data.

After describing causal discovery and identifying a method suitable for astrophysical problems involving hundreds of thousands of objects, we illustrate the technique with galaxy data.
Specifically we take seven variables describing the first-order properties (brightness, mass, size, morphology, star formation rate) of $\sim4.5\times10^5$ low-redshift galaxies from the \emph{Nasa Sloan Atlas} (NSA). We assess reliability of the causal discovery and calibrate hyperparameters of our algorithm on mock data, then infer the causal links present in the NSA including the presence of confounding variables not in the data subset.
We show explicitly how this can be used to refine our understanding of galaxy evolution and test proposed physical mechanisms, which crucially (as is typical in physics) come with a direction of operation in addition to simply inducing correlation among galaxy properties.

\section{Observational data}
\label{sec:data}

We base our analysis on the NSA v1.0.1~\citep{NSA},\footnote{\url{www.sdss4.org/dr17/manga/manga-target-selection/nsa/}} a value-added catalogue of nearby galaxies that includes inferred quantities (such as stellar mass and star formation rate) in addition to raw observables. It is based primarily on Sloan Digital Sky Survey (SDSS) imaging but employs reprocessed reductions with improved sky subtraction and photometry tailored for extended low-redshift sources.
The catalogue includes galaxies with redshifts $z\lesssim0.15$, and provides homogenised multi-band photometry and spectroscopic redshifts, including
a cross-match with Galaxy Evolution Explorer (GALEX) data to fill in the ultra-violet part of galaxies' spectral energy distributions. The NSA is a widely-studied standard for
the local galaxy population (e.g.~\citealt{nsa_4, nsa_2, nsa_1,MaNGA, nsa_3}).

We take the following fields:
\begin{itemize}
	\item \textsc{zdist}: estimated cosmological redshift correcting the observed redshift with a peculiar velocity estimate from~\citet{Willick_1997}.
	This corresponds approximately to distance in Mpc/$h$.
	\item \textsc{elpetro\_absmag}: absolute magnitude (luminosity) in the SDSS $r$-band.
	\item \textsc{elpetro\_b300}: current star formation rate (SFR) divided by the average over the past 300 Myr. This indicates how recently a galaxy formed most of its stars since 300 Myr ago.
	\item \textsc{elpetro\_mass}: Stellar mass in $M_\odot/h^2$.
	\item \textsc{sersic\_n}: S\'ersic index $n$ from a two-dimensional, single-component S\'ersic fit in the $r$-band. This indicates morphology, with $n=1$ describing an exponential disk (extremely late-type) and $n=4$ a de Vaucouleurs profile (early-type).
	\item \textsc{elpetro\_ba}: Axis ratio $b/a$ at the isophotal contour enclosing 90 per cent of a galaxy's light. This also indicates morphology, although affected by projection effects differently to $n$: low $b/a$ indicates a thin, edge-on disk, while high $b/a$ indicates a spheroid or highly inclined galaxy.
	\item \textsc{elpetro\_th50\_r}: Angular radius enclosing 50 per cent of a galaxy's light in the $r$-band, in arcseconds.
    (This could be converted to a more physically meaningful absolute size, but we do not do so for this pathfinder study because the causal link that must exist between redshift and apparent size will act as a check on the method.)
\end{itemize}
These are designed to capture the most important information about galaxy structure, including mass, luminosity, size, structure and redshift. Note that they are by no means unique: different choices would reflect different user priorities and beliefs about the most causally relevant variables.
The quantities designated `\textsc{elpetro}' derive from elliptical Petrosian flux fits, which are deemed the most reliable in the catalogue. \textsc{elpetro\_mass} and \textsc{elpetro\_b300} have been $K$-corrected to rest-frame magnitudes using the \texttt{kcorrect} code~\citep{kcorrect}.
Absolute magnitudes are given with $H_0 = 100 h$ km/s/Mpc so should be read as $M - 5\log{h}$. All logarithms are base 10.

\begin{figure*}
	\centering
	\includegraphics[width=2\columnwidth]{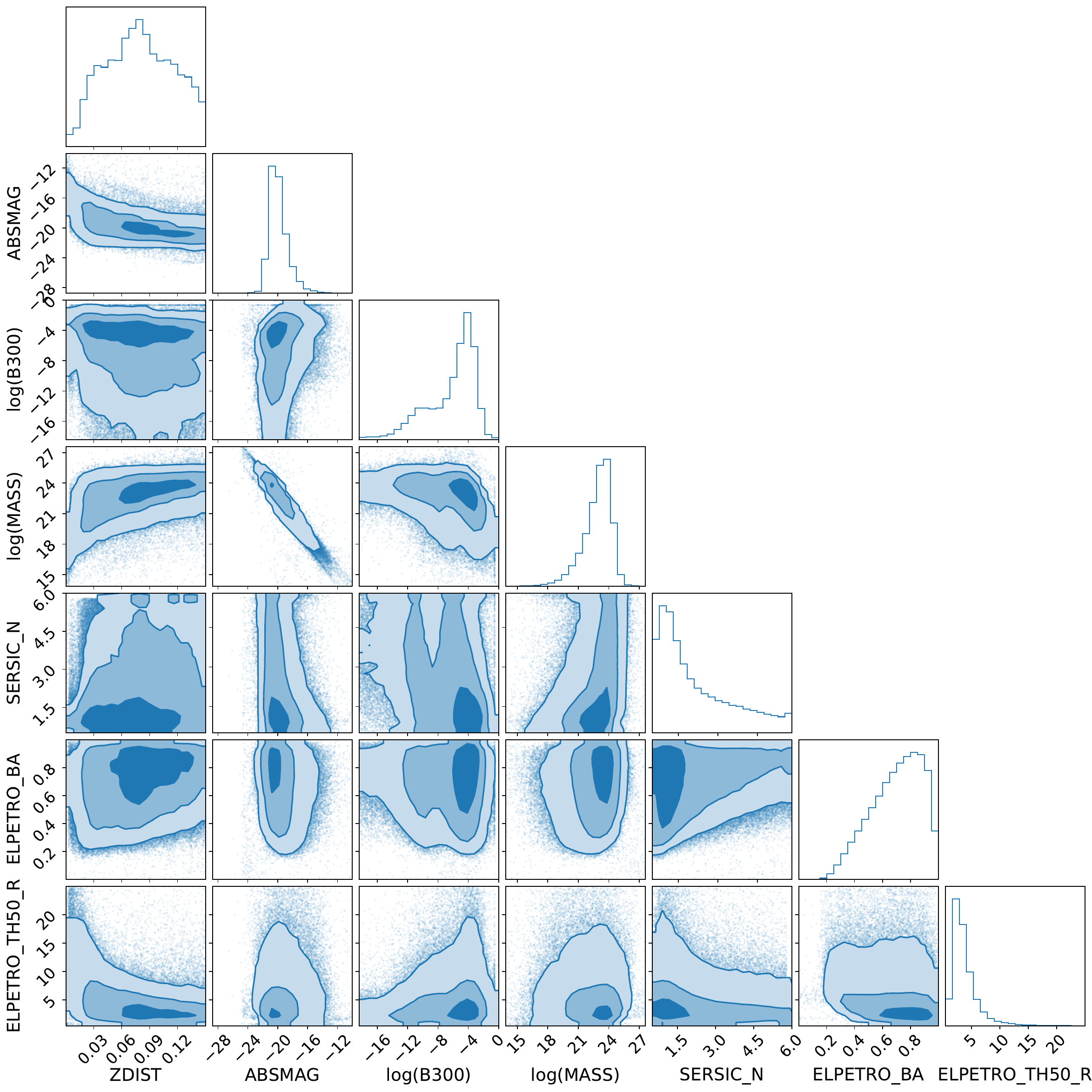}
	\caption{Distributions and pairwise correlations of the NSA data used as input to the causal discovery algorithm. The contour levels contain 39.3, 86.5 and 98.9 per cent of the points (1, 2 and 3$\sigma$ for 2D distributions). The complex correlations necessitate a nonlinear correlation metric for assessing conditional independence.}
	\label{fig:corner}
\end{figure*}

To clean the catalogue we require \textsc{zdist} $<0.15$, \textsc{elpetro\_absmag} $<-10$, \textsc{elpetro\_b300} $> 10^{-8}$, \textsc{elpetro\_b300} $< 10$, \textsc{elpetro\_mass} $> 10^6$, \textsc{elpetro\_mass} $< 10^{12}$, \textsc{sersic\_n} $< 5.8$, \textsc{elpetro\_ba} $> 0$, \textsc{elpetro\_ba} $< 1$, \textsc{elpetro\_th50\_r} $> 0$ and \textsc{elpetro\_th50\_r} $< 25$. These cuts remove anomalous objects whose properties are likely to have been inaccurately determined.
This leaves 445,763 out of an original 641,409 galaxies. A corner plot of the final dataset is shown in Fig.~\ref{fig:corner}.

\section{Methodology}
\label{sec:methods}

\subsection{Causal Discovery}
\label{sec:causal_discovery}

Most statistical analyses in astrophysics (whether or not through the lens of machine learning)
are designed to measure correlations: how strongly two quantities co-vary and the properties of their relationship.
Correlation, however, is agnostic about direction and mechanism, which are the predictions of galaxy formation theories and hence the most useful features for testing them. Projecting these predictions onto the space of correlations loses information and hence constraining power. Causal discovery methods seek to retain this information, going beyond correlation by inferring the causal structure that generates the observed data. This provides added value to the results that complements or is overlaid upon the traditional results of astrostatistical methodology.
It is this information that connects directly to theories and models of galaxy formation and evolution.
For thorough reviews of causal discovery see~\citet{review_4,review_1,review_2,review_3}.

\begin{figure}
	\centering
	\includegraphics[width=\columnwidth]{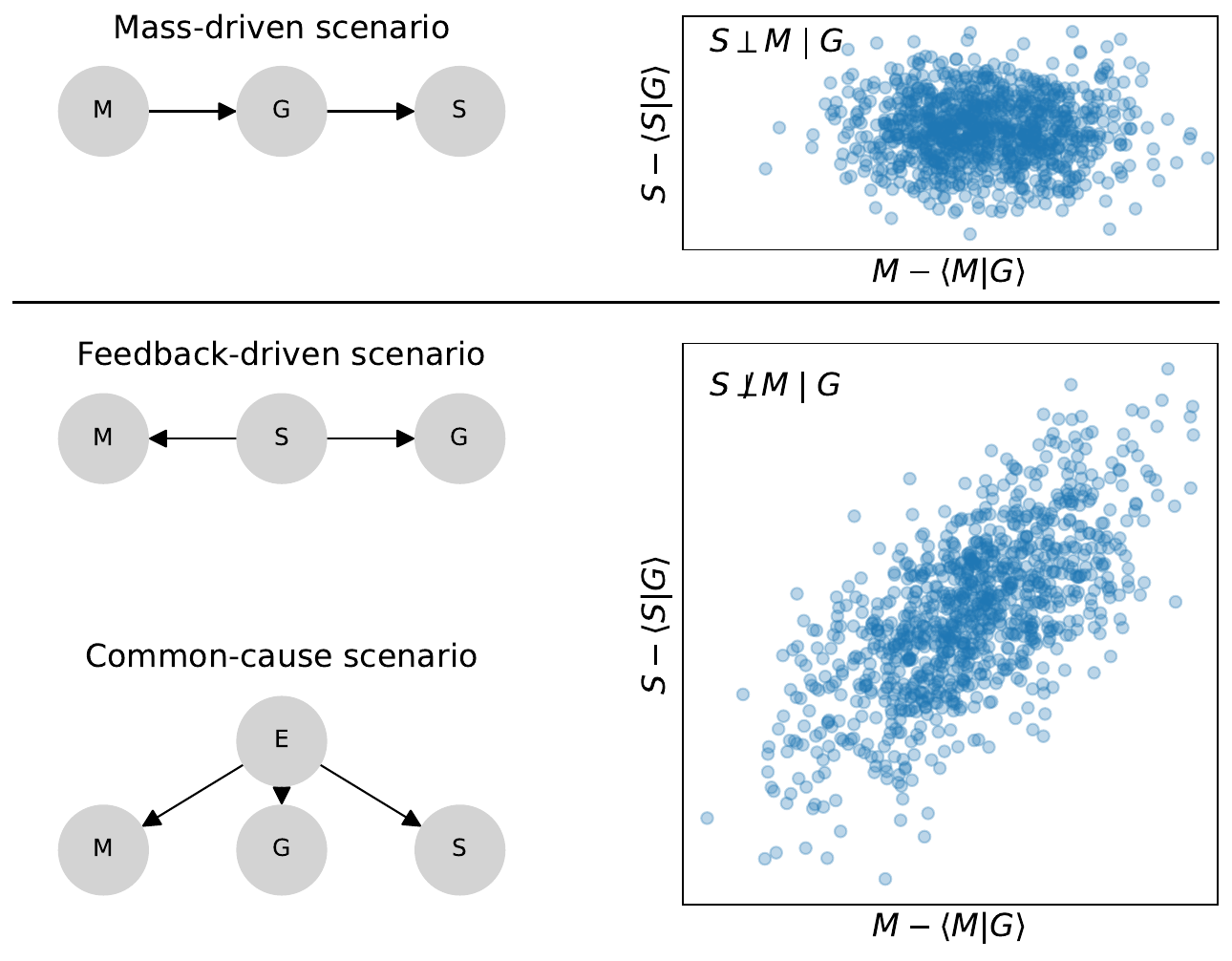}
	\caption{A visual representation of the example of Sec.~\ref{sec:causal_discovery} illustrating how causal discovery can distinguish between physical mechanisms beyond mere correlations. The axes of the scatter plots show the residuals of $S$ and $M$ upon subtraction of their expectation given their $G$ values and the average $G-S$ and $G-M$ relations. The upper causal graph has a conditional independence between $S$ and $M$ at fixed $G$, while the lower two graphs do not---the presence or absence of this conditional independence therefore narrows down the possible causal structure of the system.}
	\label{fig:example}
\end{figure}

Causal discovery utilises \emph{conditional} correlation strengths to uncover the directions of influence among variables.
To see how this works, suppose we measure three galaxy properties: total dynamical mass enclosed within the extent of the galaxy ($M$, mainly driven by dark matter), gas mass ($G$), and star-formation rate ($S$). We find all three are correlated: larger values of one are associated with larger values of any other.
From this alone, one could imagine a mass-driven scenario (more massive galaxies accrete more gas, $M\rightarrow G$, which in turn fuels star formation, $G \rightarrow S$), a feedback-driven scenario (high $S$ regulates gas supply, $S \rightarrow G$, while simultaneously drawing dark matter in through adiabatic contraction and hence increasing $M$, $S \rightarrow M$;~\citealt{Blumenthal}) or a common-cause scenario (the environment controls both mass growth and gas supply, indirectly correlating all three).
However, if $M$ and $S$ are decorrelated by conditioning on $G$ then $G$ must be responsible for linking them, which in the context of the above trichotomy implies the mass-driven scenario. Conversely, if $M$ and $S$ remain correlated even after conditioning on $G$, there must either exist a direct  causal link between them (the feedback-driven scenario), or a latent variable links them irrespective of $G$ (the common-cause scenario). This is illustrated in Fig.~\ref{fig:example}.
Thus the existence or absence of the conditional correlation breaks the degeneracy between the physical scenarios.

By assessing all such conditional correlations
(including a multi-dimensional conditioning set)
one can determine the \emph{Markov equivalence class} to which the data belongs~\citep{Verma_1990}.
Each class comprises causal structures sharing the same statistical dependencies which therefore cannot be distinguished without experimental intervention, impossible in astrophysics.
(Such statistical independencies can also be thought of as implying factorisability of the joint (probability) distribution describing the variables: e.g. $A \rightarrow B \rightarrow C$ implies $P(A,B,C) = P(A)P(B|A)P(C|B)$.) This leads to the classic \emph{constraint-based} causal discovery method, the Peter--Clark algorithm~\citep{Peter_Clark}, which eliminates \emph{direct} (i.e. causal) correlations with conditional independence tests then applies orientation rules to fill in directions where possible.\footnote{As an example of such a rule, suppose that one has identified the direct links $A-B$, $A-C$, $B-D$, $C-D$. If $B$ and $C$ are disconnected when conditioning on $A$ but not when conditioning on $D$, it must be that neither $B$ nor $C$ are caused by $D$. Hence the $B-D$ and $C-D$ links must be $B\rightarrow D$ and $C\rightarrow D$. There are two more rules involving collisions that can be used to determine directions.}
Alternative \emph{score-based} methods such Greedy Equivalence Search~\citep{Chickering_2003} assign likelihoods to candidate solutions based on the correlation strengths and search for the highest-scoring structures, while functional causal models such as additive noise models instead exploit asymmetries in functional relationships to determine causal direction.


Causal structure is visualised using a \emph{causal graph}, a graphical representation in which nodes correspond to observed variables and edges encode direct causal relationships. The true data-generating process is described by a \emph{directed acyclic graph} (DAG), in which all edges are directed (represented by an arrow) and causal cycles are forbidden \citep{Pearl_2009}. A DAG encodes both causal relations and a set of conditional independence constraints via the \emph{Markov property}: two variables are independent given a conditioning set if all paths connecting them in the causal graph are blocked by the conditioned variables, where causal colliders along the path (two variables mutually causing a third) do not have conditioned descendants. This implies that there is no link between them through variables outside the conditioning set, an important constraint on their causal relation~{\citep{Lauritzen1996}}. However, when only observational data are available the full DAG is typically not identifiable and only the Markov equivalence class can be determined. This is represented by a \emph{completed partially directed acyclic graph} (CPDAG), in which some edges are oriented while others remain undirected, indicating causal directions that cannot be uniquely determined from conditional independencies alone~\citep{Andersson,Chickering_2003}.


{The CPDAG representation relies on the assumption of \emph{causal sufficiency}, namely that all common causes of the observed variables are themselves observed. When this assumption is violated, latent (i.e. unobserved) confounding variables may induce statistical dependencies that cannot be explained by any DAG over the observed variables alone~{\citep{review_4}}. In such cases, every possible orientation of edges among observables leads to inconsistencies with the observed conditional independence structure. To accommodate this, the DAG over all variables (observed and latent) can be projected onto the observed variables, yielding a \emph{maximal ancestral graph} (MAG;~\citealt{Richardson}). A MAG allows for both directed edges as in a DAG or CPDAG, representing direct causal influence, and bidirected edges ($\leftrightarrow$) , representing the presence of an unobserved common cause.}

{As with DAGs, MAGs are generally not uniquely identifiable from observational data because the conditional independence structure does not pick out a specific graph. The equivalence-class representation in this setting is the \emph{partial ancestral graph} (PAG), which summarises all MAGs consistent with the observed conditional independences~\citep{Zhang}. In a PAG, edge endpoints may be marked with arrowheads, tails, or circles. An arrowhead ($\rightarrow$) indicates that the endpoint cannot be a cause of the adjacent variable in any compatible MAG, while a tail ($-$) indicates that it must be an cause. A circle endpoint ($\circ$) denotes ambiguity, meaning the corresponding ancestral relationship cannot be resolved and may be impacted by latent confounders. For example, an edge $X\circ-\circ Y$) implies consistency with a direct causal effect between $X$ and $Y$ in either direction, a bidirected edge due to an unobserved confounder, or a combination thereof. PAGs therefore encode both causal direction uncertainty and uncertainty arising from hidden variables.
When analysing data using conditional independence information, recovering a PAG is the aim of causal discovery algorithms that do not assume causal sufficiency. This is typically crucial in real-world applications where one cannot be sure that all causally relevant variables are included in one's dataset.}

Several assumptions are required for causal discovery to be possible.
The most common are the \emph{Markov condition} (separated variables in the causal graph are statistically independent), \emph{faithfulness} (no accidental statistical independences) and \emph{acyclicity} (nothing can be indirectly {caused by itself}). In addition, the conditional correlations must be adequately captured by the statistical test applied (which come with their own assumptions) and the threshold $p$-value chosen to identify insignificant correlations. Some methods further assume \emph{causal sufficiency}, namely that all causally relevant variables are included in the dataset.
Since causal discovery cannot guarantee the ``true'' causal graph, but rather identifies the set of structures compatible with the observed data and assumptions, it is best used to weed out statistically implausible causal relationships and hence generate testable causal hypotheses to be compared with theoretical expectations or followed up with targeted observations.

\subsection{The FCIT algorithm applied to galaxies}
\label{sec:application}

For application to the NSA, we have the requirements that a method is 1) accurate in the presence of confounding latent variables, since it is highly unlikely that all relevant information is contained in the dataset, 2) able to accommodate highly non-linear functional relationships (see Fig.~\ref{fig:corner}), and 3) efficient enough to analyse $N\approx\mathcal{O}(10^5)$ objects in reasonable time.
This is impossible for traditional algorithms, which scale as $N^3$ or worse when accounting for confounders and nonlinearity and become prohibitively memory intensive with $\gtrsim500$ samples~\citep{Kalisch2007,Strobl2019,Raghu}.

To achieve this we adopt the newly-developed method \emph{Fast Causal Inference with Targeted Testing} (FCIT;~\citealt{FCIT}) as implemented in \texttt{py-tetrad}~\citep{ramsey2023py}.\footnote{\url{https://www.cmu.edu/dietrich/philosophy/tetrad/use-tetrad/tetrad-python.html}}$^,$\footnote{\url{https://github.com/cmu-phil/py-tetrad}} {This is a hybrid constraint-and-score-based algorithm designed to scale to high-dimensional settings while retaining robustness to latent confounding and nonlinear relationships.}

{FCIT proceeds in three main stages:
\begin{enumerate}
    \item \emph{Skeleton discovery.} Starting from a complete undirected graph, FCIT iteratively removes edges by testing conditional independences of increasing conditioning set size. Unlike the standard Fast Causal Inference algorithm \cite{review_4}, FCIT restricts attention to conditioning sets that are local to the endpoints of each edge, which significantly reduces the number of tests required. It also incorporates discriminating path checks during edge removal, ensuring that edges are properly oriented before deciding on conditional independence.
    \item \emph{Targeted blocking of remaining dependencies.} For pairs of variables that remain dependent after standard local conditioning, FCIT applies a targeted testing strategy based on recursive blocking. This procedure searches for conditioning sets that block all remaining inducing paths between two variables by conditioning only on noncolliders. This targeted approach avoids the combinatorial explosion associated with exhaustive conditioning-set searches.
    Additional checks guarantee that interim and final graphs are well-formed PAGs when applied to data samples: if a candidate edge removal would violate this criterion, it is rejected. This is a departure from previous latent variable procedures that yield PAGs only theoretically.
    \item \emph{Orientation and PAG construction.} Once the adjacency structure is determined, FCIT applies orientation rules analogous to those used in FCI to identify unshielded colliders and propagate edge orientations, resulting in a PAG that encodes both causal directionality and ambiguity due to latent confounding.
\end{enumerate}}

{By combining localised conditioning with targeted testing, FCIT achieves substantial computational savings relative to FCI while empirically retaining strong recovery of PAG structure in both simulated and real data settings.}
The resulting graphs are edge-minimal and correctly oriented,
with an unprecedented runtime of only $\sim$1 minute for $4.5\times10^5$ datapoints and 7 features. {High accuracy on causal discovery benchmarks is demonstrated in~\cite{FCIT}.}

Our independence test is \texttt{use\_basis\_function\_lrt}.
{This implements a conditional independence test based on truncated basis-function expansions of the conditional mean functions. Specifically, to test for $X\perp Y\ |\ Z$ the variables $X$ and $Y$ are each expanded in an orthogonal basis (Legendre polynomials in our implementation), truncated at order given by the user-supplied hyperparameter \texttt{truncation\_limit}. Conditional independence is then assessed by comparing a restricted model in which the basis coefficients of $X$ do not enter the regression for $Y$ to an unrestricted model in which they do using a likelihood ratio test, i.e. using the null and alternative hypotheses
\begin{equation}
    H_0: Y = f(Z) + \epsilon \; \; ;\; \; \;\;H_1: Y = f(Z) + g(X) + \epsilon
\end{equation}
where $f$ and $g$ are basis-function expansions and $\epsilon$ a Gaussian noise. Independence is accepted if the improvement in log-likelihood allowing for the dependence on $X$ fails to exceed the $\chi^2$ threshold at the chosen significance level. This allows for flexible nonlinear dependencies while retaining a well-defined parametric test statistic.}

For scoring we adopt \texttt{use\_basis\_function\_bic},
{which defines a likelihood by assuming additive noise models with finite variance after basis expansion. Specifically, the likelihood $\mathcal{L}$ is computed from the corresponding sums of squares of the residuals and therefore assumes that the noise term simply adds onto the dependent variable in each regression. Model complexity is penalised by the number of retained basis coefficients using the Bayesian information criterion (BIC;~\citealt{BIC}),
\begin{equation}\label{eq:bic}
	\text{BIC} = \mathcal{L} - \texttt{penalty\_discount} \times k \ln(N),
\end{equation}
where $N$ is the sample size used for the local regression and $k$ is the number of free regression coefficients in the basis‑function expansion for the child given its current parent set~\citep{CPN}.
The BIC score balances goodness-of-fit against model complexity, with higher scores indicating preferred models in our convention. These assumptions are weaker than linear-Gaussian models while remaining computationally tractable at large sample sizes. The score-based search itself is done with the Best Order Score Search (BOSS; \citealt{BOSS}) algorithm.}

This makes FCIT a hybrid algorithm: the constraint phase uses the basis‑function likelihood ratio test to prune edges, while the score phase (and the refinement and orientation phases) uses the basis‑function BIC.
For the $p$-value threshold we use 0.01 in all cases, which we find not to affect the results appreciably. {Note that FCIT is agnostic to the specific conditional independence test and score employed. We exploit this flexibility by combining FCIT with basis-function likelihood ratio tests and scores, allowing nonlinear dependencies to be detected without sacrificing scalability.}

Two main hyperparameters affecting dataset-specific performance: \texttt{penalty\_discount} and \texttt{truncation\_limit}. The former controls how strongly graph complexity is penalised in the BIC score (Eq.~\ref{eq:bic}). A higher value favours simpler graphs by removing more noise-dominated edges, at the expense of the quality of the conditional fit as described by $\mathcal{L}$.
The latter controls the complexity of the local regression model through the number of polynomial basis terms included {in the \texttt{use\_basis\_function\_lrt} test as described above}. Larger values allow more expressive models at the cost of runtime and enhanced BIC complexity penalty.

\subsection{Mock data generation}
\label{sec:mocks}

To optimise these hyperparameters for our astrophysical application we create mock datasets with similar characteristics to the real data but with known causal structure. This will also enable the reliability of the method to be quantified, giving an indication of the systematic uncertainty possible in the results on the real data.

This is achieved with the Causal Perceptron Network (CPN;~\citealt{CPN}), a simulation framework for generating
synthetic datasets from arbitrary nonlinear models. The user specifies a DAG that encodes the desired causal structure, along with a noise distribution. Each variable is then expressed as a nonlinear function of its causes plus an independent noise term. Rather than choosing simple algebraic forms for these functions CPN uses randomly configured multilayer perceptrons, ensuring that the resulting data exhibits realistic nonlinear dependencies while remaining stable even with a large number of variables.

Each dataset is made by recursively sampling noise and propagating values forward through the causal graph, producing independent and identically distributed samples. {In detail, this means that for each variable $X_i$, a structural equation of the form
\begin{equation}\label{eq:mock_mean}
X_i = f_i\!\left(\mathrm{Pa}(X_i),\, E_i\right)
\end{equation}
is defined, where $\mathrm{Pa}(X_i)$ denotes the set of parent variables of $X_i$ in the DAG, $f_i$ is a feed-forward neural network, and $E_i$ is an independent exogenous noise term.
The noise variables $E_i$ are sampled independently and then the variables computed in the causal order of the DAG. This forward propagation ensures that each variable depends only on its parents and its own noise term. Because the noise enters as an explicit input to the neural network rather than being added post hoc, the resulting conditional distributions
$p(X_i \mid \mathrm{Pa}(X_i))$ can be highly nonlinear, non-Gaussian, and heteroskedastic. This lets CPN produce highly flexible data
suitable for testing causal discovery methods under realistic nonlinear data-generating processes. Note that the measured values from the real dataset are not used in this procedure.}

We generate datasets with the same size as the NSA subset (445,763 points) with 7 nodes and a random number of edges between 12 and 16, roughly matching what will be measured in the real data. This takes around a minute per dataset. We use four hidden layers with 50 neurons each and a Rectified Linear Unit (ReLU) activation function.
{The noise term $E_i$ is drawn independently for each sample from the default beta distribution Beta$(E; 2, 5)$. This is defined by
\begin{equation}
    \text{Beta}(E;\alpha,\beta) = \frac{E^{\alpha-1} (1-E)^{\beta-1} \Gamma(\alpha+\beta)}{\Gamma(\alpha)\Gamma(\beta)}
\end{equation}
where $\Gamma$ is the Gamma function.
This skewed distribution produces a realistic amount of non-Gaussianity in the scatters, generating} mock datasets with correlations visually similar to Fig.~\ref{fig:corner}. We nevertheless check explicitly that the results are not sensitive to reasonable variations in the noise {(see also Sec.~\ref{sec:res_mocks})}.

We then refit each of these datasets with the FCIT algorithm for a range of \texttt{truncation\_limit} and \texttt{penalty\_discount} values. For each one we compute the precision (fraction of predicted edges that are correct), recall (fraction of true edges that were successfully recovered) and F1 score (harmonic mean of the precision and recall) of the PAG produced.

\section{Results}
\label{sec:results}

\subsection{Mock data tests}
\label{sec:res_mocks}

We find that \texttt{truncation\_limit} $=14$ is ideal for this data: it is considerably larger than the default value of 3, allowing the highly nonlinear relations between variables to be captured in the FCIT scoring and conditional independence tests,
but still larger values tend to decrease the BIC due to the additional model complexity. The results are in any case
largely insensitive to this. Fixing this we then scan through \texttt{penalty\_discount}, calculating in each case the average precision, recall and F1 score across 200 mock datasets differing only in their number of edges and the random number generation.

\begin{figure}
	\centering
	\includegraphics[width=\columnwidth]{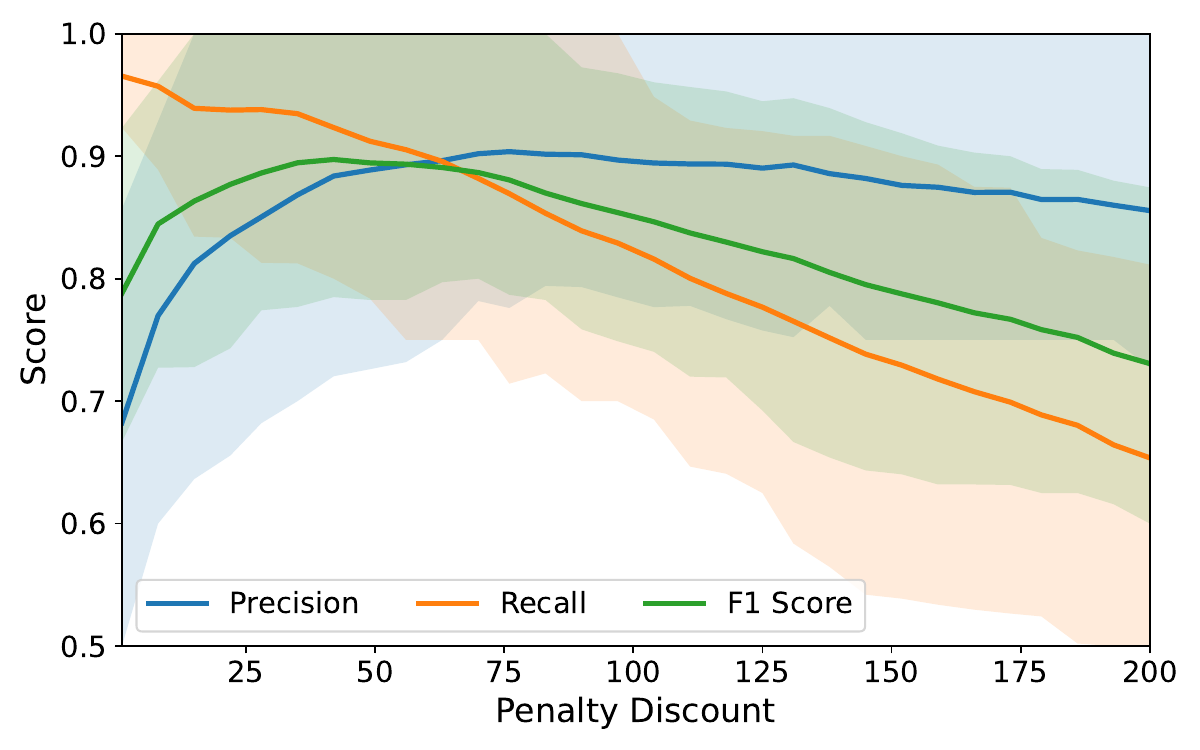}
	\caption{The precision, recall and F1 statistics across 200 NSA-like mock datasets as a function of the \texttt{penalty\_discount},
    at \texttt{truncation\_limit} $=14$. Solid lines show the mean over the datasets, and bands the 16$^\text{th}$ to 84$^\text{th}$ percentile range. A maximum reliability of $\sim$90 per cent is achieved at \texttt{penalty\_discount}$\approx$50.}
	\label{fig:mocks}
\end{figure}

The result is shown in Fig.~\ref{fig:mocks}.
As \texttt{penalty\_discount} increases the score penalises model complexity more strongly, leading to sparser graphs. This reduces false positives and thus tends to increase precision, but also causes some true edges to be missed, lowering recall. The F1 score peaks at \texttt{penalty\_discount} $\sim40-50$ at a value $\sim0.9$, roughly indicating a 90 per cent success rate on each dataset. We adopt a value of 50 for the real data.

{While the mock data matches the real data in the number of datapoints, nodes and (approximately, since this is a post-hoc statistic) causal edges, it does not match it in noise distribution, which is unknown and likely complex in the NSA. This may be expected to incur some systematic error in the optimal \texttt{trunction\_limit} and \texttt{penalty\_discount} when applied to the real data. However, the correlations induced in the mock data are qualitatively similar to those of Fig.~\ref{fig:corner}, and we explicitly check that the results of Fig.~\ref{fig:mocks} are not appreciably altered by reasonable variations in the neural network architecture or noise distribution parameters. Further, we investigate modest variations in \texttt{trunction\_limit}, \texttt{penalty\_discount} and the other hyperparameters in the FCIT algorithm and its testing and scoring methods directly on the real data, finding relatively robust results.}

\subsection{Real data}
\label{sec:res_real}

\begin{figure}
	\centering
	\includegraphics[width=0.95\columnwidth]{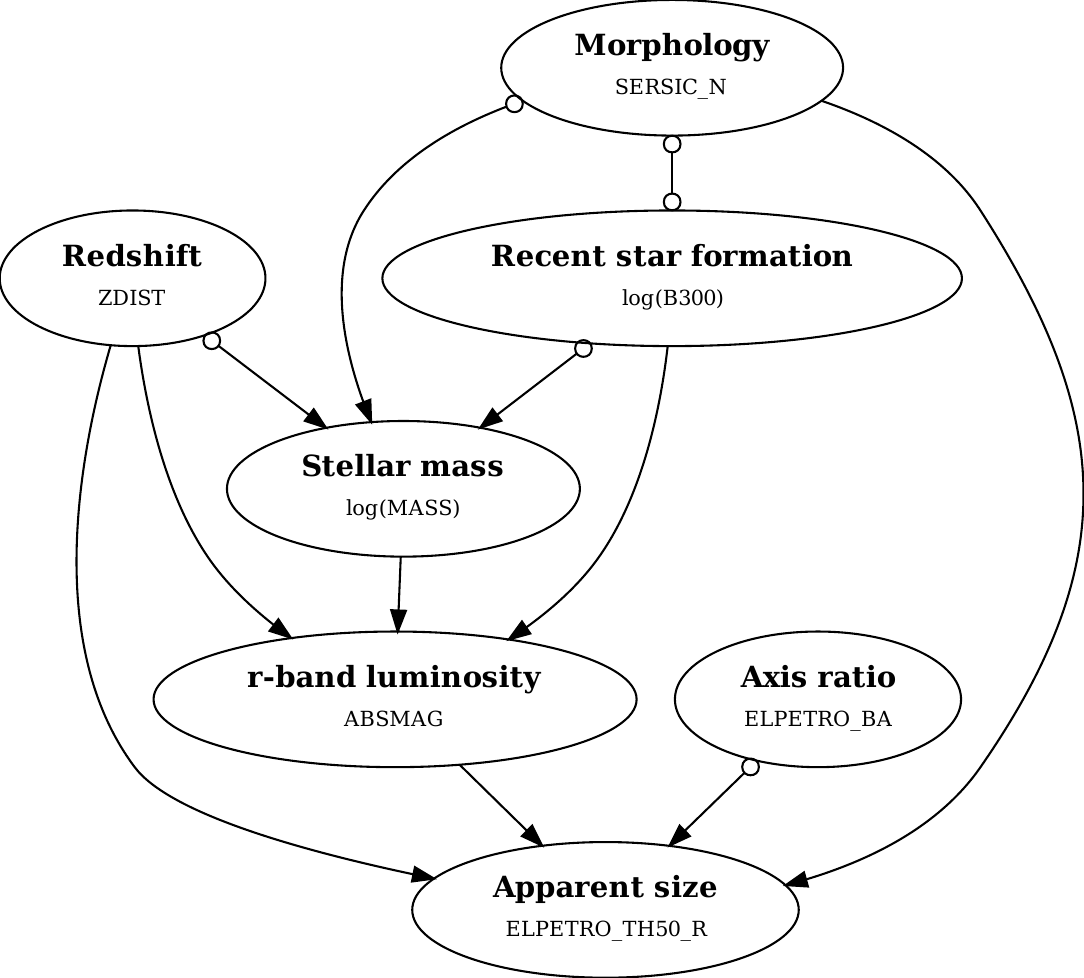}
	\caption{The PAG of the NSA data. Each node contains a colloquial parameter name in bold as well as the technical variable name in the NSA. Confident causal structures are indicated by directed edges, while less confident associations (circle endpoints) may be impacted by latent confounders.}
	\label{fig:data}
\end{figure}

We now apply the FCIT algorithm to the NSA data. The PAG produced is shown in Fig.~\ref{fig:data}.

The result describes a combination of physical effects carrying information about galaxy evolution and observational and selection effects describing the way in which the data was obtained. As expected, redshift determines apparent size, which scales inversely with angular diameter distance. It also  influences mass and absolute magnitude through Malmquist bias, the  preferential detection of intrinsically brighter objects at higher distance~{\citep{Eddington1914, Malmquist_1922}. Larger stellar mass leads to higher luminosity (ABSMAG) simply because more stars produce more light. Likewise, we expect more massive (or more luminous) galaxies to be larger: in SDSS the median half-light radius grows with mass as roughly $R \propto M^{0.4}$ for late types and even steeper ($\sim M^{0.55}$) for early types \citep{Shen2003}. Our finding ABSMAG $\rightarrow$ ELPETRO\_TH50\_R is consistent with these scaling laws~\citep{Courteau2007,Bernardi2014,Lange2015}, but crucially exposes directional causal information.
Late-type (disk) galaxies follow a shallow size--mass relation ($R_{\rm eff} \propto M_*^{0.22}$ for $M_* \gtrsim 3\times10^{9}\,M_\odot$), whereas early types follow a steeper relation ($R_{\rm eff} \propto M_*^{0.75}$) \citep{vanderWel2014}. The arrow from luminosity to size therefore reflects the well-known mass--size relation in galaxy surveys \citep{Shen2003,vanderWel2014}, although it is interesting that it is luminosity rather than mass that appears to be the fundamental driver.}

{The less obvious edges, and the graph structure as a whole, may be interpreted as follows:
\begin{itemize}
\item \emph{Size growth and disk assembly.}
The absence of a direct edge between recent star formation (log(B300)) and size indicates no evidence that short-timescale star formation independently sets galaxy size once stellar mass, luminosity, and morphology are accounted for.
Nevertheless, the causal chain log(B300) $\rightarrow$ ABSMAG $\rightarrow$ ELPETRO\_TH50\_R remains physically meaningful,
implying inside-out disk growth as a plausible physical interpretation rather than a directly inferred causal link~\citep{MunozMateos2011,Nelson2016}.
Observationally, galaxies with extended, actively star-forming outer disks exhibit larger blue-band radii and colour gradients consistent with gradual disk build-up at large radii \citep{Wang2011}.
Overall, the PAG disfavours models in which instantaneous star formation efficiency alone drives size growth, instead supporting scenarios where size is primarily regulated by longer-timescale mass assembly and structural evolution, with star formation acting as a secondary or correlated process~\citep{Brook2012,Pilkington2012}.
\item \emph{Merger-driven size evolution.}
The directed influence of luminosity (and implicitly mass) on size is also naturally interpreted in terms of merger-driven growth.
Massive early-type galaxies are known to experience substantial size evolution through minor and major mergers at late times \citep{vanDokkum2010,LopezSanjuan2012,Hilz2013}.
Since $z \sim 1$, massive ellipticals have undergone on average $\sim 0.9$ mergers, sufficient to explain $\sim 50$--75 per cent of their observed increase in half-light radius \citep{LopezSanjuan2012}.
The recovered luminosity-to-size pathway is consistent with this hierarchical picture, in which mass assembly through mergers increases galaxy size more efficiently than stellar mass alone.
\item \emph{Mass, luminosity, and star formation.}
The edge log(MASS) $\rightarrow$ ABSMAG reflects the fundamental mass--luminosity relation.
The additional edge log(B300) $\rightarrow$ ABSMAG captures the contribution of young stellar populations to optical luminosity, with recent star formation temporarily brightening galaxies.
Together, these edges indicate that luminosity encodes both long-term mass assembly and short-term star formation activity.
The partially directed connection between star formation and mass is consistent with the star-forming main sequence, ${\rm SFR} \propto M_*^{\alpha}$ with $\alpha \sim 0.6$ at low redshift \citep{Whitaker2012,Speagle2014}, while allowing for latent confounding by gas supply or environment~\citep{Peng2010,Tacconi2018}.
\item \emph{Morphology as a mediator.}
The S\'ersic index (SERSIC\_N) plays a central role in the PAG, with a directed edge to size and uncertain connections to both stellar mass and star formation.
As a proxy for bulge dominance, $n \simeq 1$ traces disk-dominated systems, while $n \simeq 4$ corresponds to bulge-dominated galaxies \citep{GrahamDriver2005,Conselice2014}.
The edge SERSIC\_N $\rightarrow$ ELPETRO\_TH50\_R indicates that morphology causally influences size at fixed luminosity, consistent with the fact that bulge-dominated systems are more compact. Here ``causal'' should be understood in the structural sense of the PAG: SERSIC\_N acts as a proxy for the cumulative outcome of physical processes such as merging, disk instabilities, and quenching, rather than as a fundamental dynamical variable~\citep{KormendyKennicutt2004}.
The circle endpoints on edges linking S\'ersic index to mass and star formation rate indicate uncertainty in causal direction or the presence of latent variables, such as merger history or environment, that jointly affect morphology, mass growth, and quenching.
\item \emph{Uncertain edges and latent confounding.}
Several further edges appear with circle endpoints, particularly among morphology, mass, and star formation.
These likely reflect the influence of unobserved confounders including halo mass, gas accretion history, feedback efficiency, and large-scale environment.
Similarly, ambiguous links involving the apparent axis ratio (ELPETRO\_BA) may arise from inclination-dependent dust attenuation, which affects apparent luminosity and measured structural parameters.
In the PAG formalism, such circle endpoints explicitly encode the limits of identifiability given the conditional independence structure of the available variables and sample size.
\item \emph{Synthesis with galaxy-formation theory.}
Overall, the PAG supports a picture in which stellar mass assembly underlies galaxy evolution, setting luminosity and, through it, galaxy size~\citep{NaabOstriker2017}.
Morphology acts as a key mediator, reflecting the cumulative impact of mergers and internal dynamical processes, while recent star formation modulates luminosity without independently determining size.
The recovered structure is consistent with observational evidence that, at fixed mass, late-type galaxies are larger than early types and that galaxy sizes evolve as $R_{\rm eff} \propto (1+z)^{-0.75}$ for disks and $\sim (1+z)^{-1.48}$ for ellipticals \citep{vanderWel2014}.
The PAG therefore captures well-established empirical regularities and distinguishing between plausible physical pathways that would otherwise be empirically degenerate, while clearly delineating where the data is not sufficient to support stronger causal claims. This is highly promising for the future, where more sophisticated applications of the methodology to data of higher quality and quantity may yield crucial insight into the physical mechanisms of galaxy formation.
\end{itemize}
}

\section{Discussion and Conclusion}
\label{sec:conc}

The application of causal discovery to astrophysics is largely virgin territory. By enabling the directions of physical links to be established, it provides a significant information overlay on (even machine learning-based) correlation analyses, helping to constrain theories that postulate the physical mechanisms governing the data. Such theories essentially correspond to DAGs, so causal discovery can be seen as a method for inferring theories (as far as is possible) directly from data.

To illustrate the approach we have applied causal discovery to low-redshift galaxy data from the NSA, adapting a hybrid constraint-and-score algorithm---FCIT---to meet the demands of astrophysical data (large datasets, highly nonlinear correlations and presence of confounders). After testing and calibrating the method on NSA-like mock data (establishing $\sim$90 per cent accuracy) we applied it to the real data to find the PAG in Fig.~\ref{fig:data}. This supports a hierarchical, {morphology- and} mass-driven framework of galaxy evolution while indicating the complexities involved in the physical mechanisms at play. It also highlights the vital importance of \emph{observational} causal discovery methods, since intervention is impossible in astrophysics. Such methods are sometimes undervalued in favour of intervention-based methods involving randomised control trials.

In the near term there are several ways in which this analysis could be extended. First, many of the causal links in Fig.~\ref{fig:data} reflect observational or selection effects rather than physical mechanisms. The data could be refined to minimise these, for example by conditioning properties on redshift or constructing combinations of variables less prone to selection biases or trivial correlations. Second, the several ambiguous (circle) endpoints indicate the potential impact of latent variables not included in the dataset. By folding in such properties as gas mass, metallicity, dust attenuation and environmental density these ambiguities could be resolved, providing a clearer picture of the overall flow of causality. There is of course a huge range of further data across astrophysics that could profitably be interpreted through a causal discovery lens.

There is room for improvement on the theoretical side too. While we showed good performance, our method still relies on choices of conditional independence tests and scoring which have not been explored exhaustively. Mismatches between the mock and real data could lead to errors being larger in the latter. Besides directly investigating other hyperparameter choices, mock datasets and algorithms, this could be investigated by applying causal discovery to cosmological simulations, which have known physical mechanisms but more accurately capture correlations likely to be present in astrophysical data.
This would also provide a platform for investigating the loss of information due to selection effects, reveal more clearly the causal graphs associated with candidate physical models and calibrate how reliably different orientations can be distinguished. In parallel, bootstrap and stability analyses on the observational data could determine which edges are most robust.

General and efficient causal discovery methods remain under active development. Future algorithms might be able to resolve circle endpoints to distinguish models within a PAG equivalence class by utilising other types of information (e.g. nonlinearity, non-Gaussianity or heteroskedasticity), as has already been done when assuming causal sufficiency in the linear, non-Gaussian regime (e.g.~{\citealt{Hyvarinen2010,Lingam}}). One could even do Bayesian model comparison between competing simulations or theories based on their causal structures they predict~\citep{Bayesian_CD,astro_eg_0}, directly demonstrating the gain in constraining power afforded by causal discovery.
Taken together, these steps would provide a more complete and reliable causal map, ultimately allowing causal discovery to move beyond recovering broad backbones to testing detailed models of how galaxies grow and evolve.

In summary, this study paves the way for causal discovery to become as mainstream in astrophysics as it is in other data-rich fields where causal correlations---and their directions---encode crucial information about the underlying mechanisms.

\section*{Data availability}

The Nasa Sloan Atlas is publicly available at \url{https://www.sdss4.org/dr17/manga/manga-target-selection/nsa/}.
The code used in this paper is publicly available on Github \githublink.

\section*{Acknowledgements}

We thank David Bacon, Deaglan Bartlett, Robin Evans, Pedro Ferreira, Sebastian von Hausegger, Matt Jarvis, Richard Stiskalek, Simon White and Tariq Yasin for useful discussions.
HD is supported by a Royal Society University Research Fellowship (grant no. 211046).

\bibliographystyle{mnras}
\bibliography{refs}

\end{document}